\documentclass[10pt,letterpaper]{article}
\usepackage[top=0.85in,left=2.75in,footskip=0.75in]{geometry}

\usepackage{amsmath,amssymb}

\usepackage{changepage}
\usepackage{algorithm}
\usepackage{textcomp,marvosym}

\usepackage{cite}

\usepackage{nameref,hyperref}

\usepackage[right]{lineno}

\usepackage[nopatch=eqnum]{microtype}
\DisableLigatures[f]{encoding = *, family = * }

\usepackage[table]{xcolor}

\usepackage{array}

\newcolumntype{+}{!{\vrule width 2pt}}

\newlength\savedwidth

\raggedright
\usepackage[aboveskip=1pt,labelfont=bf,labelsep=period,justification=raggedright,singlelinecheck=off]{caption}
\renewcommand{\figurename}{Fig}

\makeatletter
\renewcommand{\@biblabel}[1]{\quad#1.}
\makeatother

\usepackage{lastpage,fancyhdr,graphicx}
\usepackage{epstopdf}
\fancyheadoffset[L]{2.25in}
\fancyfootoffset[L]{2.25in}
\begin{document}
\vspace*{0.2in}

\begin{flushleft}
{\Large
\textbf\newline{Noise reduction in ISAR imaging of UAVs using weighted atomic norm minimization and 2D-ADMM algorithm} 
}
\newline
\\
Mohammad Roueinfar\textsuperscript*,
Mohammad Hossein Kahaei\textsuperscript{1},

\bigskip
\textbf{1} School of Electrical Engineering, Iran University of Science and Technology, Tehran, Iran
\\
\bigskip

%
%
%
%
%


* mrooein@hotmail.com

\end{flushleft}
\section*{Abstract}
The effect of noise on the Inverse Synthetic Aperture Radar (ISAR) with sparse apertures is a challenging issue for image reconstruction with high resolution at low Signal-to-Noise Ratios (SNRs). It is well-known that the image resolution is affected  by the bandwidth of the transmitted signal and the Coherent Processing Interval (CPI) in two dimensions, range and azimuth, respectively. To reduce the noise effect and thus  increase the two-dimensional resolution of Unmanned Aerial Vehicles (UAVs) images, we propose the Fast Reweighted Atomic Norm Denoising (FRAND) algorithm by incorporating the weighted atomic norm minimization. To solve the problem, the Two-Dimensional Alternating Direction Method of Multipliers (2D-ADMM) algorithm is developed to speed up the implementation procedure. Assuming sparse apertures for ISAR images of UAVs, we compare the proposed method with the MUltiple SIgnal Classification (MUSIC), Cadzow, and $\mathrm{SL}_{0}$ methods in different SNRs. Simulation results show the superiority of FRAND at low SNRs based on the Mean-Square Error (MSE), Peak Signal-to-Noise ratio (PSNR) and Structural Similarity Index Measure (SSIM) criteria.


\nolinenumbers

\section{Introduction}
The use of Unmanned Aerial Vehicle (UAV) has expanded significantly in various applications, such as monitoring forest fires, firefighting, aerial photography, product delivery, transportation, agriculture, traffic control, infrastructure inspection, etc. ~\cite{bib1}, ~\cite{bib2}, ~\cite{bib3}, ~\cite{bib4}. Along with progressive research on this area, there are also some concerns about the unauthorized use of UAVs, which pose threats such as invasion of privacy of people and places, and destruction  ~\cite{bib5}, which motivates more investigation on this area more seriously. To do so, different methods using sound, heat, optics, and radar have been presented. However, due to the low flight altitude, low heat, and weak sound of UAVs, the detection procedure becomes more difficult especially when the distance between the UAV and the detector is long~\cite{bib6}.
In ~\cite{bib7}, long-wave infrared thermal radiation is used for UAV detection at night. In general, optical methods are more suitable for the detection and classification of small UAVs at short ranges, but they suffer from low or variable lights and also obstacles effects. Also, sound detection methods used for the direction of arrival-based identification are effectively appropriate for short distances,  and distinguishing small UAVs like quadcopters from birds ~\cite{bib8}. Radio or passive methods that use the downlink signal of quadcopters for detection may increase false alarms due to interference ~\cite{bib9}.

The Laser Imaging Detection and Ranging (LIDAR) technique, which operates by emitting laser pulses, is very sensitive to weather conditions, smoke, and direct sunlight ~\cite{bib8}. In contrast, radar-based methods are more effective for detecting quadcopters ~\cite{bib8}, as they are immune to light and adverse weather conditions. For example, in ~\cite{bib10}, an Frequency Modulation Continuous Wave (FMCW) method is used to detect the micro-Doppler pattern caused by the UAV rotor. In ~\cite{bib11},  the impact  of frequency, polarization, and other parameters on ISAR images is investigated using the radar signature of small drones.  Also, an FMCW radar network has been developed by ~\cite{bib8} to detect and classify multi-rotor UAVs.

In some applications, high-resolution imaging of UAVs is required under various weather conditions.  An effective method applied for aerial targets is the Inverse Synthetic Aperture Radar (ISAR)~\cite{bib12}, which produces two-dimensional images of moving targets in the range and azimuth directions. These images are created by the echo signals of target scatterers received from different angles and also processing the Doppler frequency caused by the target motion. {Using  the squint minimization technique in FMCW ISAR, the large trajectory deviations of UAVs are reduced, ~\cite{bib13}. In ~\cite{bib14} for accurate detection and positioning of UAV swrams, a method based on the Dechirp-keystone transform and frequency-selective reweighted trace minimization is developed.
A tracking-before-detection algorithm is reported in ~\cite{bib15} for UAV tracking  based on multiple ellipses or a sub-random matrices model with non-linear ISAR observations and unknown orientation angle due to the Doppler effect.
In ~\cite{bib16}, a method  is presented for UAV imaging based on an improved complex variable mode decomposition approach and using a microwave  photonic ISAR.

As known, the amplitude of scattered signals  is related to the target Radar Cross Section (RCS) ~\cite{bib17}, and the resolution in range and azimuth directions depends on the transmit signal bandwidth and the CPI, respectively. Effectively, the wider the bandwidth, the more resolution in the range direction is achieved. In practice, however, the bandwidth and CPI are limited and can not be increased to the desired amounts.
From another aspect, it is well known that for retrieving ISAR images, scatterer points with higher RCS are the most effective and generally sparse. Moreover, some radar pulses may be missed during transmission/reception, or due to maneuvers of the target leading to  sparse apertures, which can result in blurred images ~\cite{bib18}. To improve the ISAR image resolution in such situations, we can exploit the target sparsity and  sparse apertures in the spatial domain. 

This leads to  super-resolution methods, which are divided into on-grid, off-grid, and gridless methods. In the first case, we consider a set of discrete fixed points and assume the target scatterers are located on them. Then, the resolution is improved for a larger number of grids, which in turn increases the complexity.
In the off-grid method, the distances between the pre-determined grids and scatterers are estimated simultaneously and included in computations, which normally results in a complicated implementation. The gridless method, also known as the continuous compressed sensing method, is defined using Atomic Norm Minimization (ANM), where, unlike the previous two methods, a set of continuous atoms is defined based on the received signal (echo). It is assumed that the echoes of target scatterers are linearly
combined, where each echo is represented by an atom that is a complex exponential vector whose amplitude and phase are proportional to the RCS and phase of each echo, respectively. By applying convex minimization to the problem, the atoms, target scatterers, and finally the target image are obtained. The gridless method with ANM has been applied for ISAR imaging in  ~\cite{bib19}, ~\cite{bib20}, and ~\cite{bib21}. Also, ~\cite{bib22}, ~\cite{bib23}, and ~\cite{bib24}  have presented some methods to reduce the noise effect at low SNRs.

In this paper,  we propose the FRAND method by incorporating the weighted atomic norm to reduce the noise effect and increase the resolution in ISAR imaging. The paper is organized as follows. In Section 2, the signal model is defined. The FRAND method is introduced in Section 3 and the corresponding simulation results are presented in Section 4. Section 5 concludes the paper.

\section{Signal Model}
The relative motion between a radar and target results in translational and rotational motions, which lead to blurring and range-variant phase errors in ISAR imaging, respectively ~\cite{bib25}.   In this work, we assume both translational and rotational motions are compensated. Also, we use the Frequency Stepped Chirp Signal (FSCS) shown in Fig~\ref{fig1}. This signal is active in L specific and equal time intervals ($L$ bursts), where each burst consists of $M$ pulses with Linear Frequency Modulation (LFM), $B_{1}$ shows the frequency bandwidth of each pulse, $\Delta f$ is the size of the frequency step, $T_{1}$ is the pulse duration, $M$ is the number of frequency steps, and $\frac{1}{T}$ determines the Pulse Repetition Frequency (PRF).
\begin{figure}
\begin{adjustwidth}{-1.2in}{0in}
\includegraphics[angle=0,width=1.3\textwidth]{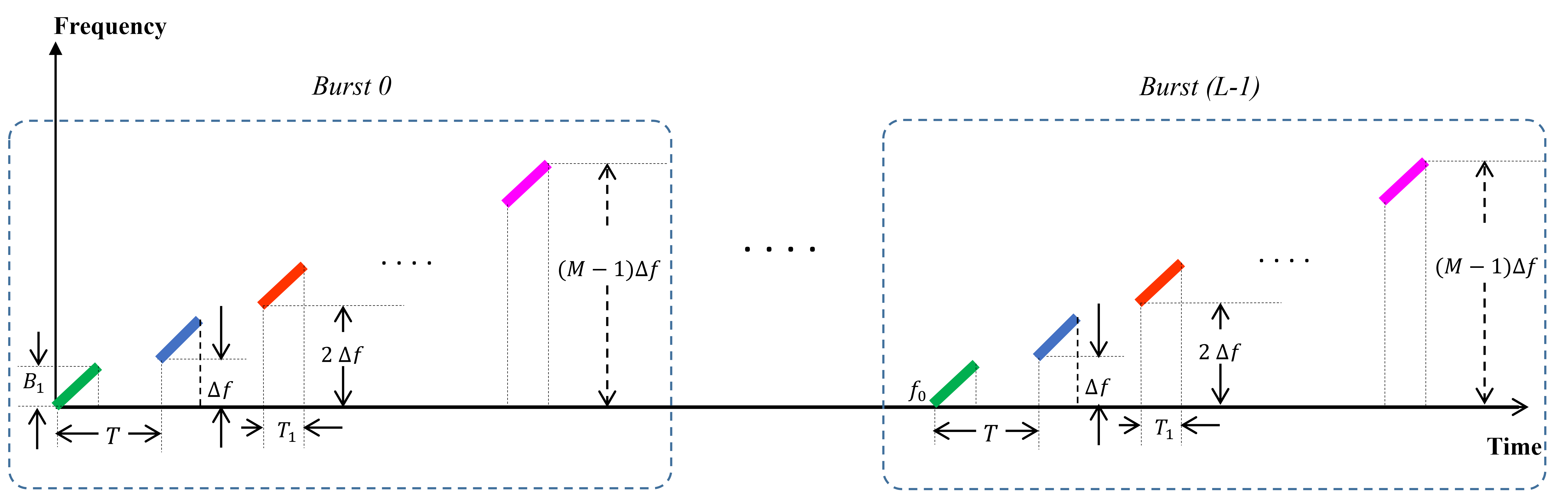}
\caption{\bf FSCS time-frequency diagram.}
\label{fig1}
\end{adjustwidth}
\end{figure}
It is well-known in pulse compression radars that the range resolution is inversely proportional to the transmitted signal bandwidth. In the FSCS structure, the bandwidth of an LFM is split into some narrowband sub-chirps, and after properly processing this waveform with the echoes,  the bandwidth of the composite signal at the matched filter output effectively is increased, which consequently improves the range resolution. Also, due to the use of narrowband sub-chirps in this waveform,  other radar transmitters have the least interference.\\
 
Mathematically, the $\textit{m}$-th pulse of the $\textit{l}$-th burst for $l=0,1,...,L-1$ and $m=0,1,...,M-1 $ is defined as
\begin{eqnarray}
\label{eq:e1}
s_{\mathrm{T}}(t,m;l) =x(t-(lM+m)T) \exp[\mathrm{j}2\pi f_{\mathrm{m}}(t-(lM+m)T)],
\end{eqnarray}
where $x(t)=rect(\frac{t}{T_{1}}) \exp[\mathrm{j}\pi \gamma t^{2}]$, $T_{1}$ is the pulse width,  $\gamma= \frac{\Delta f}{T_{1}}$  is the chirp rate, $f_{\mathrm{m}}=f_{0}+m \Delta f$ is the carrier frequency of the $\textit{m}$-th pulse with $f_{0}$ showing the initial frequency, and $\Delta f$ is the frequency step. Accordingly, the received signal composed of $K$ scatterers is described as
\begin{align}
\label{eq:e2}
 r_{\mathrm{T}}(& t,m;l) = \nonumber \\
& \sum_{k=0}^{K-1}  \sigma_{k}x(t-\tau_{k}(t)-(lM+m)T) \exp[\mathrm{j}2\pi f_{\mathrm{m}}(t-\tau_{k}(t)-(lM+m)T)],
\end{align}
where $\sigma_k$ is the reflection coefficient, $\tau_{k}(t)=\frac{2R_{k} (t_{\mathrm{n}})}{c}$ is the delay it takes the ISAR signal to travel to the $\textit{k}$-th scatterer and return, with $R_k (t_{\mathrm{n}})$ being the instantaneous distance  at the discrete observation time, $t_{\mathrm{n}}$, $n=0,1,...,N-1$,  in the azimuth direction for the signal received from angle $\theta_{\mathrm{n}}$.
Then, after dechirping, the received signal is obtained as
\begin{eqnarray}
\label{eq:e3}
r(f_{\mathrm{m}},t_{\mathrm{n}}) =\sum_{k=0}^{K-1} \sigma_{k}\exp \left[-\mathrm{j}2\pi f_{\mathrm{m}}(\frac{2R_{k} (t_{\mathrm{n}}))}{c}\right].
\end{eqnarray}
By associating the rotational motion of the target at the coordinates $(x_{k}, y_{k})$, the instantaneous range of ISAR to the center of the $\textit{k}$-th scatterer is measured as
\begin{eqnarray}
\label{eq:e4}
R_{k} (t_{\mathrm{n}})= \sqrt{R_{0}^{2}+x_{k}^{2}+y_{k}^{2}+2R_{0} [x_{k}  \sin(\omega t_{\mathrm{n}} )+y_{k}  \cos(\omega t_{\mathrm{n}} ) ], }
\end{eqnarray}
where $\omega$ is the angular velocity and $R_{0}$ shows the distance between the center of the target and radar. Since this distance is usually large; that is the target is in far field, we can approximate Eq(\ref{eq:e4}) as
\begin{eqnarray}
\label{eq:e5}
\quad R_{k} (t_{\mathrm{n}} )\approx R_{0}+x_{k}  \sin(\omega t_{\mathrm{n}} )+y_{k}  \cos(\omega t_{\mathrm{n}} ),
\end{eqnarray}
which can be accordingly presented as
\begin{eqnarray}
\label{eq:e6}
\quad R_{k} (t_{\mathrm{n}} )\approx R_{0}+x_{k}  \sin(\theta_{\mathrm{n}} )+y_{k}  \cos(\theta_{\mathrm{n}} ),
\end{eqnarray}
with $\theta_{\mathrm{n}}=\omega t_{\mathrm{n}}$ showing the observation angle at $t_{\mathrm{n}}$.

By substituting Eq(\ref{eq:e6}) in Eq(\ref{eq:e3}), the received signal model can be written as
\begin{eqnarray}
\label{eq:e8}
r(f_{\mathrm{m}}, \theta_{\mathrm{n}}) = \exp\left[-\mathrm{j}4\pi f_{\mathrm{m}} \frac{R_{0}}{c}\right] \sum_{k=0}^{K-1} \sigma_k \exp\left[-\mathrm{j}4\pi f_{\mathrm{m}} \frac{x_{k} \theta_{\mathrm{n}}}{c}\right] \exp\left[-\mathrm{j}4\pi f_{\mathrm{m}} \frac{y_{k}}{c}\right],
\end{eqnarray}
where we assumed the rotation angle $\theta_{\mathrm{n}}$ is typically small to use the approximations $\sin(\theta_{\mathrm{n}}) \approx \theta_{\mathrm{n}}$ and $\cos(\theta_{\mathrm{n}}) \approx 1$. Furthermore, to compensate for the translational motion, we multiply  Eq(\ref{eq:e8}) by $\exp\left[\mathrm{j}4\pi f_{\mathrm{m}} \frac{R_{0}}{c}\right]$ to model the received signal at  the $\textit{n}$-th observation angle and $\textit{m}$-th pulse as
\begin{eqnarray}
\label{eq:e10}
r(f_{\mathrm{m}}, \theta_{\mathrm{n}}) = \sum_{k=0}^{K-1} \sigma_{k} \exp\left[-\mathrm{j}4\pi f_{\mathrm{m}} \frac{x_{k} \theta_{\mathrm{n}}}{c}\right] \exp\left[-\mathrm{j}4\pi f_{\mathrm{m}} \frac{y_{k}}{c}\right].
\end{eqnarray}

\section{Proposed FRAND method}
We present a method to reduce the noise effect and enhance the resolution of ISAR images.
To do so, we first define the following parameters:
\begin{eqnarray}
\label{eq:e11}
h_{\mathrm{n}}= \frac{4\pi f_{\mathrm{m}}}{c} \theta_{\mathrm{n}}\approx \frac{4\pi f_{0}}{c} \theta_{\mathrm{n}}  \qquad \mathrm{and}  \qquad
h_{\mathrm{m}}= \frac{4\pi f_{\mathrm{m}}}{c}
\end{eqnarray}
where  $f_{\mathrm{m}}=f_{0}+m\Delta f$ and the imposed approximation is acceptable since $\theta_{\mathrm{n}}$'s are small. By substituting Eq(\ref{eq:e11})  in Eq(\ref{eq:e10}), the received signal reduces to
\begin{eqnarray}
\label{eq:e13}
r(\theta_{\mathrm{n}}, f_{\mathrm{m}})=\sum_{k=0}^{K-1}\sigma_{k} \exp[-jh_{\mathrm{n}} x_{k}-jh_{\mathrm{m}} y_{k}].
\end{eqnarray}
To present Eq(\ref{eq:e13}) in matrix form, we introduce $\boldsymbol{\mathrm{R}} \in \mathbb{C} ^{N \times M}$ with the entries $[\boldsymbol{\mathrm{R}}]_{n,m}=r(\theta_{\mathrm{n}}, f_{\mathrm{m}})$ and   the matrix of atoms for the $\textit{k}$-th scatterer $\boldsymbol{\mathrm{A}}(k) \in \mathbb{C} ^{N \times M}$ with the entries
\begin{eqnarray}
\label{eq:e14}
\big[\boldsymbol{\mathrm{A}}(k) \big]_{n,m}=\exp[-jh_{\mathrm{n}} x_{k}-jh_{\mathrm{m}} y_{k}].
\end{eqnarray}
Also, we define the atomic set as
\begin{eqnarray}
\label{eq:e16}
\mathcal{A}=\Big \{\boldsymbol{\mathrm{A}}(k) \in \mathbb{C} ^{N \times M} \Big \arrowvert \quad  \big | \big[\boldsymbol{\mathrm{A}}(k) \big]_{n,m}  \big | =1 \Big\}.
\end{eqnarray}
Then, by vectorizing $\boldsymbol{\mathrm{R}}$  and   $\boldsymbol{\mathrm{A}}(k)$ as $\boldsymbol{r}=\mathrm{vec}(\boldsymbol{\mathrm{R}}) \in \mathbb{C} ^{NM \times 1} $ and $\boldsymbol{a}(k)=\mathrm{vec}(\boldsymbol{\mathrm{A}}(k)) \in \mathbb{C} ^{NM \times 1} $, respectively, we get
\begin{eqnarray}
\label{eq:e15}
\boldsymbol{r}=\sum_{k=0}^{K-1}\sigma_{k}  \boldsymbol{a}(k),
\end{eqnarray}
 In this way, according to ~\cite{bib20},  we can express the atomic norm problem  conditioned on Eq(\ref{eq:e15}) as
\begin{align}
\label{eq:e17}
\| \boldsymbol{r} \| _{\mathcal{A}}  &=  \inf \big\{t>0  : \boldsymbol{r} \in t
\mathrm{conv}(\mathcal{A}) \big\}  \nonumber \\
& = \inf_{\sigma_{k}} \Bigg\{ \sum_{k=0}^{K-1}\sigma_{k} \Bigg \arrowvert \boldsymbol{r}=\sum_{k=0}^{K-1}\sigma_{k}  \boldsymbol{a}(k) , \quad  \sigma_{k} >0  \Bigg\}.
\end{align}

where $\mathrm{conv}(\mathcal{A})$ is the convex hull of $\mathcal{A}$ . However, to estimate $\boldsymbol{r}$, note that in practice we may not receive some observations, i.e., some pulses from some angles. As a result, we consider $\boldsymbol{r}$ as the vector of complete observations and $\boldsymbol{r}_{\Omega}$ as the vector of incomplete observations caused by the sparse apertures in ISAR imaging, where $\Omega \subset \big\{ 0,1,...,(NM-1) \big\} \times 1 $ is the subset of indices of all possible observations. As a result, we obtain the estimates under the latter set as
\begin{eqnarray}
\label{eq:e18}
\boldsymbol{\hat{\boldsymbol{r}}}   =   \arg \min_{\boldsymbol{r}} \| \boldsymbol{r} \| _{\mathcal{A}}    \qquad  s.t.    \qquad  \boldsymbol{\hat{r}}_{\Omega}=\boldsymbol{\boldsymbol{r}}_{\Omega},
\end{eqnarray}
which is inherently related to the observation angles and pulse frequencies in the signal model defined by Eq(\ref{eq:e11}) to Eq(\ref{eq:e15}). By solving this problem, the lost observation angles or frequencies of unobserved pulses are effectively recovered, which can lead to increasing ISAR image resolution at low SNRs. To simplify the solution of Eq(\ref{eq:e18}), we apply theorem 1 in  ~\cite{bib20} as
\begin{align}
\label{eq:e19}
\| \boldsymbol{r} \| _{\mathcal{A}} &= \arg \min_{(\boldsymbol{u}, t)}  \Bigg\{ \frac{1}{2} \mathrm{Tr}(\boldsymbol{\mathcal{T}}(\boldsymbol{u}))+\frac{1}{2} t \Bigg\}  \nonumber \\
 & s.t.  \qquad
\begin{bmatrix}
\boldsymbol{\mathcal{T}}(\boldsymbol{u}) & \boldsymbol{r}\\
\boldsymbol{r}^{*} &  t
\end{bmatrix} \succeq 0,  \qquad \hat{\boldsymbol{r}}_{\Omega}=\boldsymbol{r}_{\Omega},
\end{align}
where $\boldsymbol{\mathcal{T}}(\boldsymbol{u}) $ is a hermitian Toeplitz matrix whose first column is the vector $\boldsymbol{u} \in  \mathbb{C} ^{N}$, $t=\sum_{k=0}^{K-1}\sigma_{k}$, and $\mathrm{Tr}(.)$ is the trace of a matrix.

To consider a practical scenario, we include Additive White Gaussian Noise (AWGN), ${\boldsymbol{n}}_{\Omega}$, as follows:
\begin{eqnarray}
\label{eq:e20}
\boldsymbol{z}_{\Omega} = \boldsymbol{r}_{\Omega}+\boldsymbol{n}_{\Omega}.
\end{eqnarray}
Based on ~\cite{bib26} and ~\cite{bib27}, to  reduce the noise effect, the following problem is defined:
\begin{eqnarray}
\label{eq:e21}
\hat{\boldsymbol{r}}   =   \arg \min_{\boldsymbol{r}} \frac{1}{2}\| \boldsymbol{r}_{\Omega}-\boldsymbol{z}_{\Omega} \| _{\mathrm{F}}^{2} +\lambda \| \boldsymbol{r} \| _{\mathcal{A}},
\end{eqnarray}
where $ \lambda $ is the regularization parameter. Next, we modify Eq(\ref{eq:e21}) by adding a weighting matrix as
\begin{eqnarray}
\label{eq:e22}
\boldsymbol{\hat{\boldsymbol{r}}}   =   \arg \min_{\boldsymbol{r}} \| \boldsymbol{r}_{\Omega}-\boldsymbol{z}_{\Omega} \| _{\mathrm{F}}^{2} +2\lambda \| \boldsymbol{\mathrm{W}}\boldsymbol{r}_{\Omega} \| _{\mathcal{A}}
\end{eqnarray}
where
\begin{eqnarray}
\label{eq:e23}
\boldsymbol{\mathrm{W}}   =  ( \tan (\boldsymbol{\mathcal{T}}(\boldsymbol{u})+\epsilon  \boldsymbol{\mathrm{I}}    )^{-1}
\end{eqnarray}
with $\epsilon$ showing an adjustment parameter. \\

According to  Eq(\ref{eq:e22}) and Eq(\ref{eq:e23}), we can rewrite Eq(\ref{eq:e19}) as
\begin{align}
\label{eq:e24}
& \arg \min_{\boldsymbol{r}} \| \boldsymbol{r}_{\Omega}-\boldsymbol{z}_{\Omega} \| _{\mathrm{F}}^{2} +2\lambda \| \boldsymbol{\mathrm{W}}\boldsymbol{r}_{\Omega} \| _{\mathcal{A}}= \nonumber \\
& \arg \min_{(\boldsymbol{u},\boldsymbol{t},\boldsymbol{r})}  \| \boldsymbol{r}_{\Omega}-\boldsymbol{z}_{\Omega} \| _{\mathrm{F}}^{2} +
2\lambda \Bigg\{\frac{1}{2} \mathrm{Tr}(\boldsymbol{\mathrm{W}} \boldsymbol{\mathcal{T}}(\boldsymbol{u}))+\frac{1}{2} t \Bigg\} \nonumber \\
&  s.t. \qquad
\begin{bmatrix}
\boldsymbol{\mathcal{T}}(\boldsymbol{u}) & \boldsymbol{r}\\
\boldsymbol{r}^{*} &  t
\end{bmatrix} \succeq 0,  \qquad \boldsymbol{\hat{\boldsymbol{r}}}_{\Omega}=\boldsymbol{r}_{\Omega}.
\end{align}
However, we should note that Eq(\ref{eq:e24}) is computationally complex with a high execution time. To cope with the latter deficiency, here we apply the 2D-ADMM algorithm. Using ~\cite{bib26} and the augmented Lagrangian,  Eq(\ref{eq:e24}) is written as
\begin{align}
\label{eq:e25}
\mathcal{L}_{\rho}(t, \boldsymbol{u}, \boldsymbol{r}, \boldsymbol{\mathrm{Z}},  \boldsymbol{\mathrm{\Lambda}}) &=
\| \boldsymbol{r}_{\Omega}-\boldsymbol{z}_{\Omega} \| _{\mathrm{F}}^{2}+\lambda \Bigg\{ \mathrm{Tr}(\boldsymbol{\boldsymbol{\mathrm{W}}\mathcal{T}}(\boldsymbol{u}))+ t \Bigg\} \nonumber \\
& + \langle \boldsymbol{{\mathrm{\Lambda}}}, \boldsymbol{\mathrm{Z}}-
\begin{bmatrix}
\boldsymbol{\mathcal{T}}(\boldsymbol{u}) & \boldsymbol{r}\\
\boldsymbol{r}^{*} &  t
\end{bmatrix}   \rangle +\frac{\rho}{2} \Bigg\| \boldsymbol{\mathrm{Z}}-
\begin{bmatrix}
\boldsymbol{\mathcal{T}}(\boldsymbol{u}) & \boldsymbol{r}\\
\boldsymbol{r}^{*} &  t
\end{bmatrix}   \Bigg\|,
\end{align}
where $\rho$ is a penalty parameter and $\boldsymbol{\Lambda}$ and $\boldsymbol{\mathrm{Z}}$ are Hermitian matrices  defined as ~\cite{bib26}
\begin{eqnarray}
\label{eq:e26}
\boldsymbol{{\mathrm{\Lambda}}}  =
\begin{bmatrix}
\boldsymbol{\mathrm{\Lambda}}_{NM \times NM} & \boldsymbol{\mathrm{\Lambda}}_{NM \times 1}\\
\boldsymbol{\mathrm{\Lambda}}_{1 \times NM} & \boldsymbol{\mathrm{\Lambda}}_{1 \times 1}
\end{bmatrix}, \qquad \mathrm{and}  \qquad
\boldsymbol{Z}  =
\begin{bmatrix}
\boldsymbol{\mathrm{Z}}_{NM \times NM} & \boldsymbol{\mathrm{Z}}_{NM \times 1}\\
\boldsymbol{\mathrm{Z}}_{1 \times NM} & \boldsymbol{\mathrm{Z}}_{1 \times 1}
\end{bmatrix}.
\end{eqnarray}
To start the recursive algorithm, we first initialize $\boldsymbol{\mathrm{\Lambda}}^{0}$ = $\boldsymbol{\mathrm{0}} $ and  $\boldsymbol{\mathrm{Z}}^{0} $ = $\boldsymbol{\mathrm{0}}$ and use the update steps as follows:
\begin{eqnarray}
\label{eq:e28}
\big\{ t^{i+1} , \boldsymbol{u}^{i+1}, \boldsymbol{r}^{i+1} \big\}  = \arg \min_{t, \boldsymbol{u}, \boldsymbol{r}} \mathcal{L}_{\rho} {(t ,\boldsymbol{u} ,\boldsymbol{r} ,  \boldsymbol{\mathrm{Z}}^{i} , \boldsymbol{\Lambda}^{i}), }
\end{eqnarray}
\begin{eqnarray}
\label{eq:e29}
\boldsymbol{\mathrm{Z}}^{i+1}   =\arg \min_{\boldsymbol{\mathrm{Z}}\succeq 0 } \mathcal{L}_{\rho} {(t^{i+1} ,\boldsymbol{u}^{i+1} ,\boldsymbol{r}^{i+1} ,  \boldsymbol{\mathrm{Z}} , \boldsymbol{\mathrm{\Lambda}}^{i}), }
\end{eqnarray}
\begin{eqnarray}
\label{eq:e30}
\boldsymbol{\mathrm{\Lambda}}^{i+1}  = \boldsymbol{\mathrm{\Lambda}}^{i}+ \rho \Bigg( \boldsymbol{\mathrm{Z}}^{i+1}-
\begin{bmatrix}
\boldsymbol{\mathrm{T'}(\boldsymbol{u}^{i+1})}  & \boldsymbol{r}^{i+1}\\
\boldsymbol{r}^{(i+1)^{H}} &   t^{l+1}
\end{bmatrix}  \Bigg),
\end{eqnarray}
where $i$ is the iteration number and  
\begin{eqnarray}
\label{eq:e31}
t^{i+1} =\frac{1}{2} \boldsymbol{\mathrm{Z}}_{1 \times 1}^{i}+\frac{1}{2} (\boldsymbol{\mathrm{Z}}_{1 \times 1}^{i})^{*} +\frac{1}{\rho} (\boldsymbol{\mathrm{\Lambda}}_{1 \times 1}^{i}-\frac{\lambda}{2}),
\end{eqnarray}
\begin{eqnarray}
\label{eq:e32}
\qquad \boldsymbol{r}^{i+1} =  \frac{1}{2\rho+1} \Bigg( \boldsymbol{\mathrm{Z}}_{0}^{i}+2\rho \boldsymbol{\mathrm{Z}}_{1}^{i}+2\boldsymbol{\mathrm{\Lambda}}_{1}^{i} \Bigg),
\end{eqnarray}
\begin{eqnarray}
\label{eq:e33}
\boldsymbol{u}^{i+1} = \boldsymbol{\mathcal{T}}^{*} \Bigg( \boldsymbol{\mathrm{Z}}_{0}^{i} +\frac{1}{\rho}\boldsymbol{\Lambda}_{0}^{i} -\frac{\lambda}{2\rho}  \Bigg).
\end{eqnarray}
According to  ~\cite{bib26}, $\boldsymbol{\mathrm{Z}}$ can be updated as
\begin{eqnarray}
\label{eq:e34}
\boldsymbol{\mathrm{Z}}^{i+1}   =\arg \min_{\boldsymbol{\mathrm{Z}}\succeq 0 } +\frac{\rho}{2} \Bigg\| \boldsymbol{\mathrm{Z}}-
\begin{bmatrix}
\boldsymbol{\mathcal{T}}^{*}(\boldsymbol{u}^{i+1}) & \boldsymbol{r}^{i+1}\\
{\boldsymbol{r}^{*}}^{i+1} &  t^{i+1}
\end{bmatrix}   \Bigg\|.
\end{eqnarray}
Finally, we  decompose $\boldsymbol{\mathrm{Z}}^{i+1}$ into eigenvalues as
\begin{eqnarray}
\label{eq:e35}
\boldsymbol{\mathrm{Z}}^{i+1}   =\boldsymbol{\mathrm{V}} \boldsymbol{\mathrm{D}} \boldsymbol{\mathrm{V}}^{*}, \quad
\end{eqnarray}
where $\boldsymbol{\mathrm{V}} $ is a square matrix whose $\textit{j}$-th column corresponds to the eigenvector of
the $\textit{j}$-th eigenvalue of $\boldsymbol{\mathrm{Z}}^{i+1}$
and  $\boldsymbol{\mathrm{D}}$ is a matrix  whose diagonal elements are the eigenvalues of $\boldsymbol{\mathrm{Z}} $. Then, according to ~\cite{bib28}, non-zero eigenvalues gives the frequencies of the atoms corresponding to the range and azimuth direction of target scatterers.
The summary of FRAND is given in Algorithm 1. \\
\begin{algorithm}[!h]
\small
\caption{ISAR imaging procedure for the proposed FRAND. }
{\textbf{Input}: Received signal $\boldsymbol{r}$ }\\
{\textbf{Output}: Recovered ISAR image of the target}
\begin{enumerate}
\small
\item Definition of atoms   $\boldsymbol{a}(k)$ for $K$ scatterers.
\item Rewriting the received signal in vector form according to Eq(\ref{eq:e15})
\item Definition of weighted atomic norm $\|\boldsymbol{r}\|_\mathcal{A}$ according to Eq(\ref{eq:e22})
\item Definition of weighting matrix according to  Eq(\ref{eq:e23})
\item Implementation of the proposed FRAND method according to Eq(\ref{eq:e24})
\item Starting the 2D-ADMM method with initial values $\boldsymbol{\mathrm{\Lambda}}^{0}$ = $\boldsymbol{\mathrm{0}}$ and $\boldsymbol{\mathrm{Z}}^{0}$ = $\boldsymbol{\mathrm{0}}$ 
\item Repeating Eq(\ref{eq:e29})
\item Updating Eq(\ref{eq:e30})
\item Updating Eq(\ref{eq:e31})
\item Updating Eq(\ref{eq:e32})
\item Decomposing $\boldsymbol{\mathrm{Z}}^{i+1}=\boldsymbol{\mathrm{V}}\boldsymbol{\mathrm{D}}\boldsymbol{\mathrm{V}}^{*}$ into eigenvalues
\item Recovering the frequencies corresponding to the target scatterers
\item Recovering the target image
\end{enumerate}
\end{algorithm}
\section{Simulation results}
We simulate the proposed FRAND method for radar imaging of a moving quadcopter. The transmit signal is an FSCS with a center frequency of 10 GHz, chirp bandwidth of 500 MHz for one burst, PRF of 6 kHz, and pulse width of 0.4$\mu$ sec. The signal is corrupted by AWGN. It is assumed that the speed of the rotor and propeller of the quadcopter is so high that their images are almost fixed. Also, the quadcopter has no rotational motion and the translational motion is assumed to be approximately linear that is compensated. The simulations are  performed on an Intel (R) Core (TM) i7-10700 CPU @ 2.90 GHz processor with 16 GB of RAM.

The performance of FRAND is compared to that of the MUSIC ~\cite{bib23}, Cadzow ~\cite{bib24},  and $\mathrm{SL}_{0}$  ~\cite{bib27}  methods.
The recovered images are shown in  Fig~\ref{fig3} for 100 to 1600 available samples  at 10 dB SNR.
As can be seen, even using 600 samples out of 1600 possible samples, the quadcopter can be recognized. This is due to the sparse apertures in  ISAR imaging, which is obtained completely randomly. Obviously, a larger number of samples can increase the image resolution with more details.
\begin{figure}[!h]
\begin{adjustwidth}{-2in}{0in}
\includegraphics[angle=0,width=1.5\textwidth]{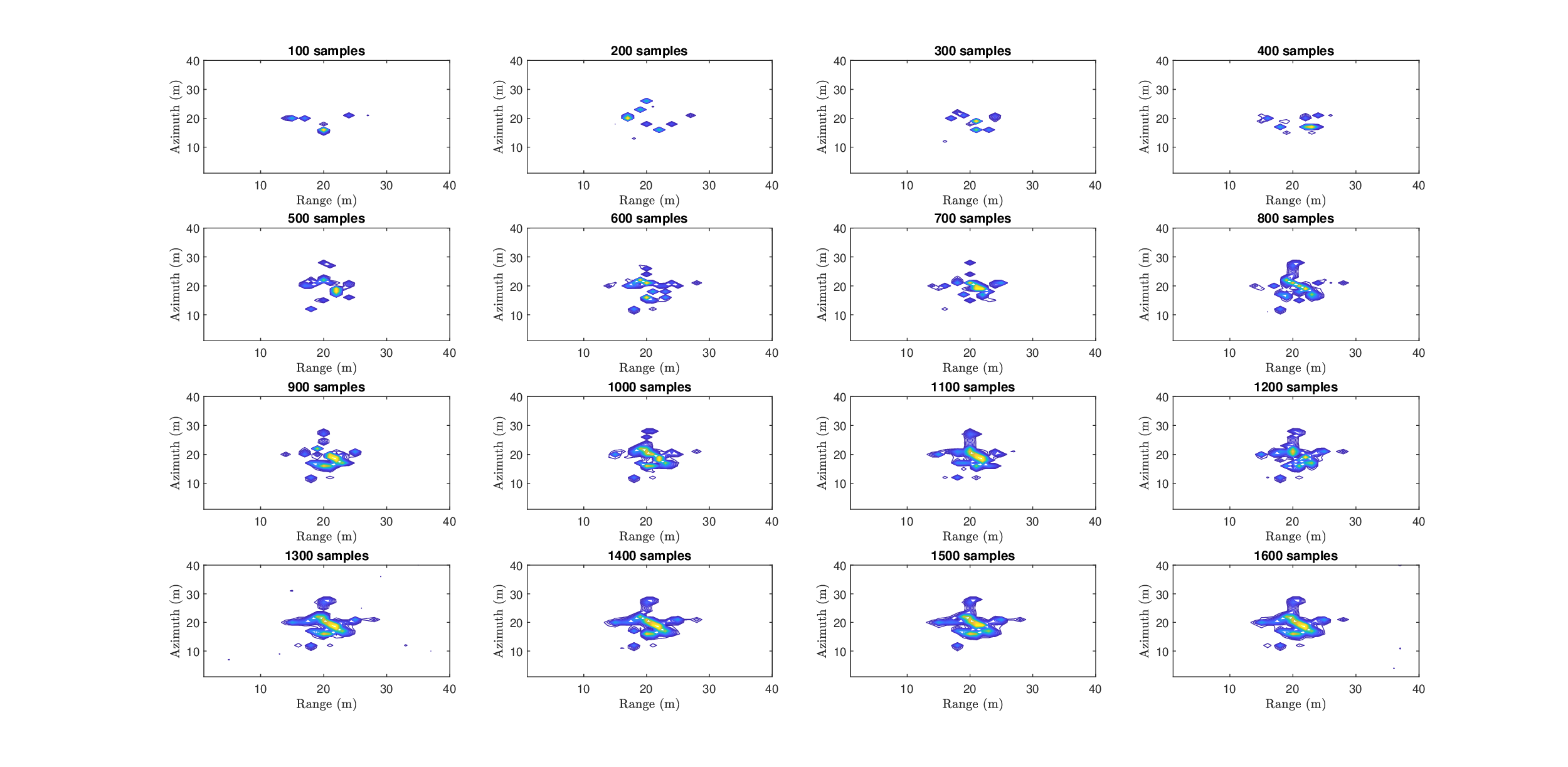}
\caption{Recovered images of a quadcopter by FRAND using 100 to 1600 available samples at 10 dB SNR. }
\label{fig3}
\end{adjustwidth}
\end{figure}

In Fig~\ref{fig4}, FRAND is compared with the mentioned methods at different SNRs using 500 samples. It is seen that FRAND and $\mathrm{SL}_{0}$ outperform the others by acceptably recovering the target image at lower SNRs; like 0 dB, due to a better noise reduction. However, when the number of samples reduces to 100, as shown in Fig~\ref{fig5}, the performance of $\mathrm{SL}_{0}$ gradually reduces at  low SNRs, while FRAND still generates recognizable images.
\begin{figure}
\begin{adjustwidth}{-2in}{0in}
\includegraphics[angle=0,width=1.5\textwidth]{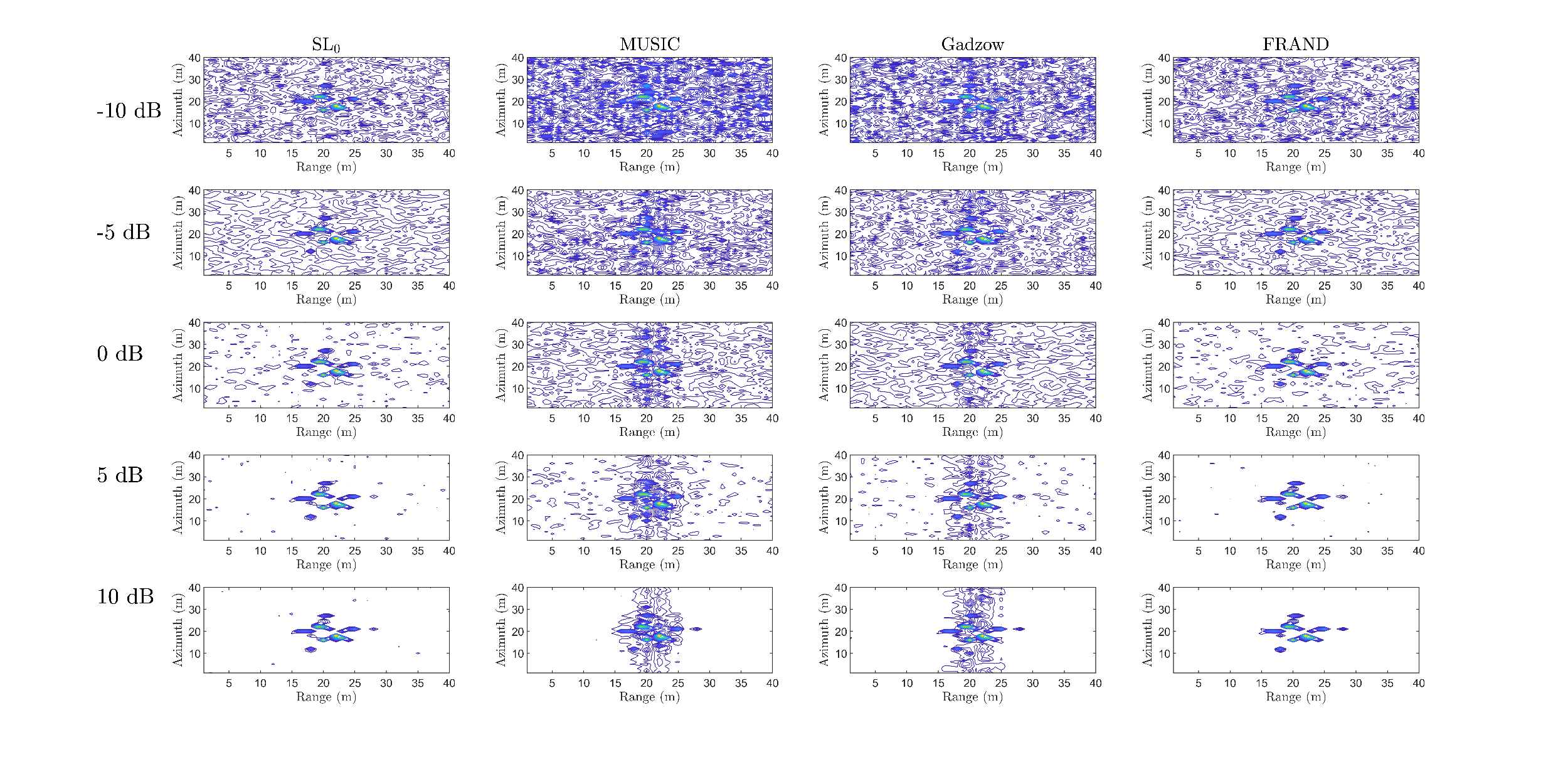}
\caption{Recovered images of a quadcopter by $\mathrm{SL}_{0}$, MUSIC, Cadzow, and FRAND for 500 available samples at SNRs= -10, -5, 0, 5, and 10 dB.}
\label{fig4}
\end{adjustwidth}
\end{figure}
\begin{figure}
\begin{adjustwidth}{-2in}{0in}
\includegraphics[angle=0,width=1.5\textwidth]{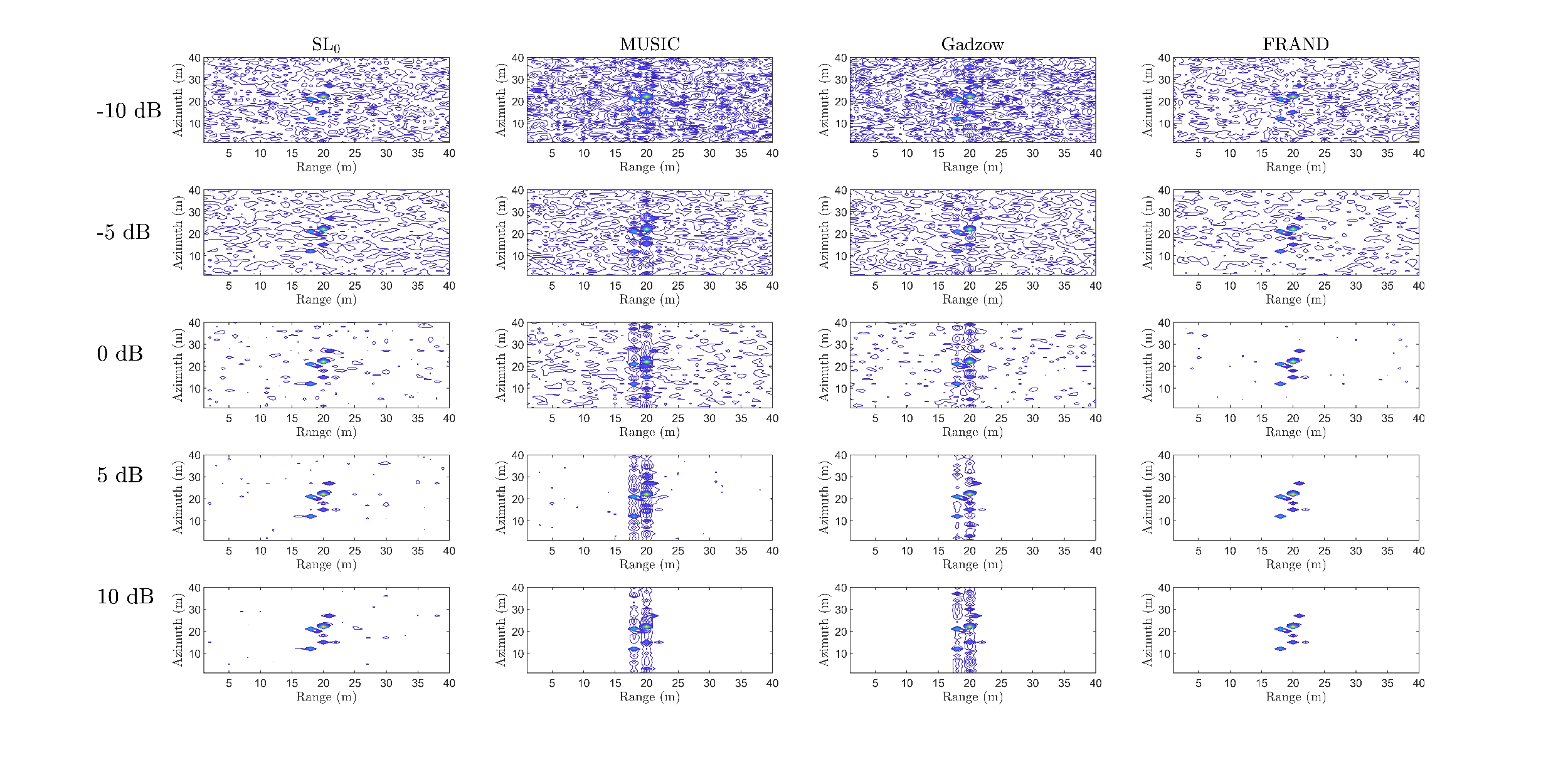}
\caption{Recovered images of the quadcopter using $\mathrm{SL}_{0}$, MUSIC, Cadzow, and FRAND for 100 available samples at SNRs= -10, -5, 0, 5, and 10 dB.}
\label{fig5}
\end{adjustwidth}
\end{figure}
Next, we assess the algorithms in terms of the MSE criterion for -10 to 10 dB SNRs. The results are shown in Fig~\ref{fig6} by averaging 100 independent trials of the experiment for each algorithm.
\begin{figure}
\includegraphics[width=0.65\textwidth]{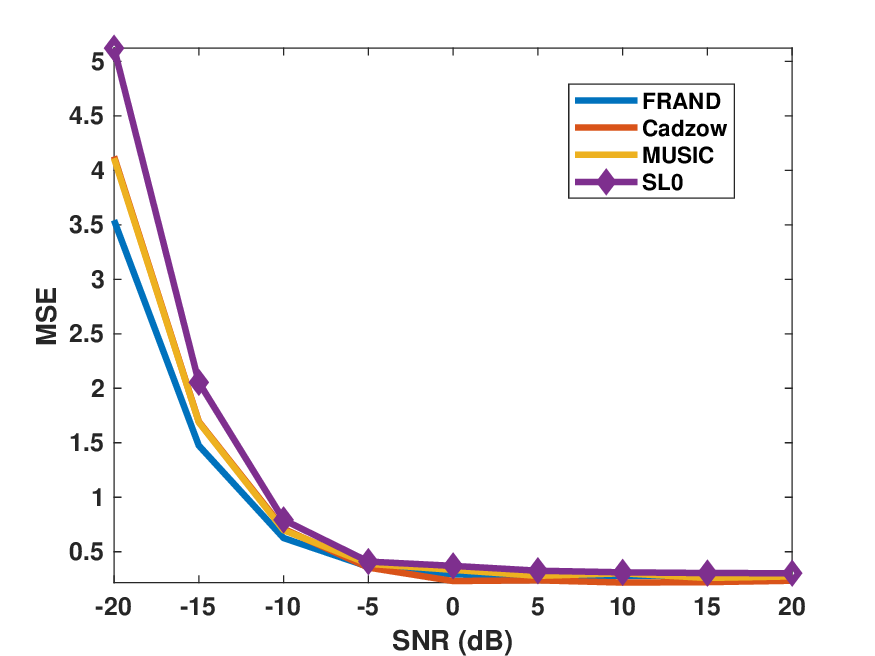}
\caption{Comparison of FRAND with MUSIC, Cadzow, and $\mathrm{SL}_{0}$ at SNRs= -10 to 10 dB in terms of MSE.}
\label{fig6}
\end{figure}
As can be seen, FRAND generates lower MSE than the other methods, especially  at lower SNRs. The numerical values of MSEs are also given in Table~\ref{Tab.1} for a detailed comparison.
\begin{table}
\begin{adjustwidth}{0in}{0in}
\caption{{\bf MSE of $\mathrm{SL}_{0}$, MUSIC, Cadzow, and FRAND at different SNRs (dB).}}
{\begin{tabular}{l   |   l    l    l   l}
\hline 
\hline 
SNR (dB)  & $\mathrm{SL}_{0}$ &	MUSIC &	Cadzow &FRAND    \\
\hline
-20  & 3.4387 &	2.6586 &	2.7280 &	2.2965   \\
\hline
-15  & 1.3786 &	1.1098 &	1.1157 &	0.9427  \\
\hline
-10  & 0.5596 &	0.4682 &	0.4626 &	0.4158  \\
\hline
-5   & 0.3316 &	0.2588 &    0.2351 &	0.2481  \\
\hline
0    &  0.3583 &	0.2153 &	0.1738 &	0.2131  \\
\hline
5    & 0.3626 &	0.2205 &	0.1637 &	0.2080   \\
\hline
10   &  0.3635 &	0.2098 &	0.1697 &    0.2072   \\
\hline
15   & 0.3634 &	0.1982 &	0.1773 &	0.2073  \\
\hline
20   & 0.3632 &	0.1925 &	0.1623 &	0.2071  \\
\hline  \hline
\end{tabular}
}
\label{Tab.1}
\end{adjustwidth}
\end{table}
These algorithms are also compared based on the Peak Signal-to-Noise ratio (PSNR) in Fig~\ref{fig7} by averaging 100 independent runs of each experiment.   One can see that the image quality recovered by FRAND is superior to the other images at the SNRs lower than -5 dB due to better noise reduction.  Moreover, the algorithms are compared in Figure  Fig~\ref{fig8} based on the Structural Similarity Index Measure (SSIM), where outperformance of FRAND is observed at different SNRs.
\begin{figure}
\includegraphics[width=0.65\textwidth]{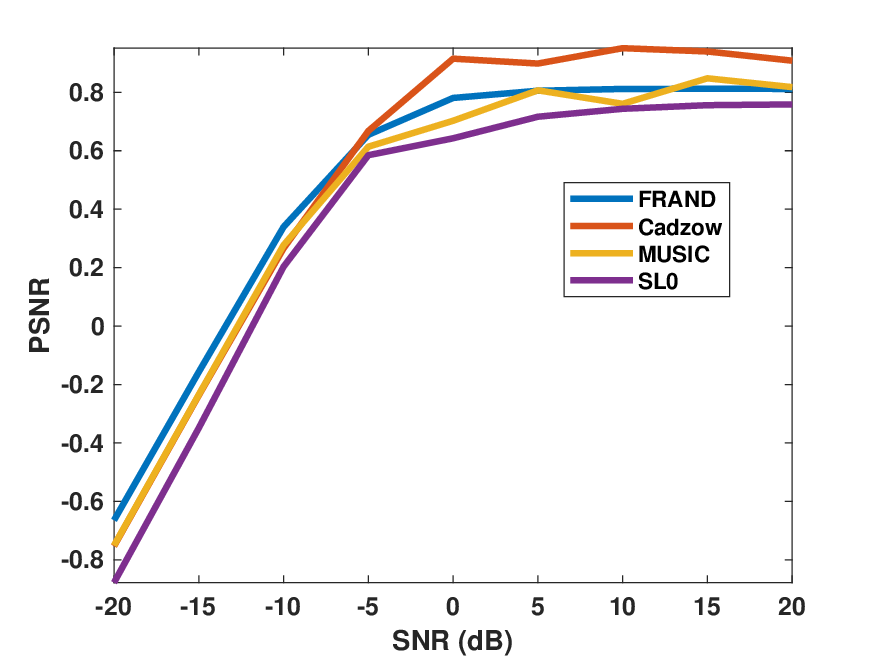}
\caption{Comparison of FRAND with MUSIC, Cadzow, and $\mathrm{SL}_{0}$ at SNRs= -10 to 10 dB in terms of PSNR.}
\label{fig7}
\end{figure}

\begin{figure}
\includegraphics[angle=0,width=0.65\textwidth]{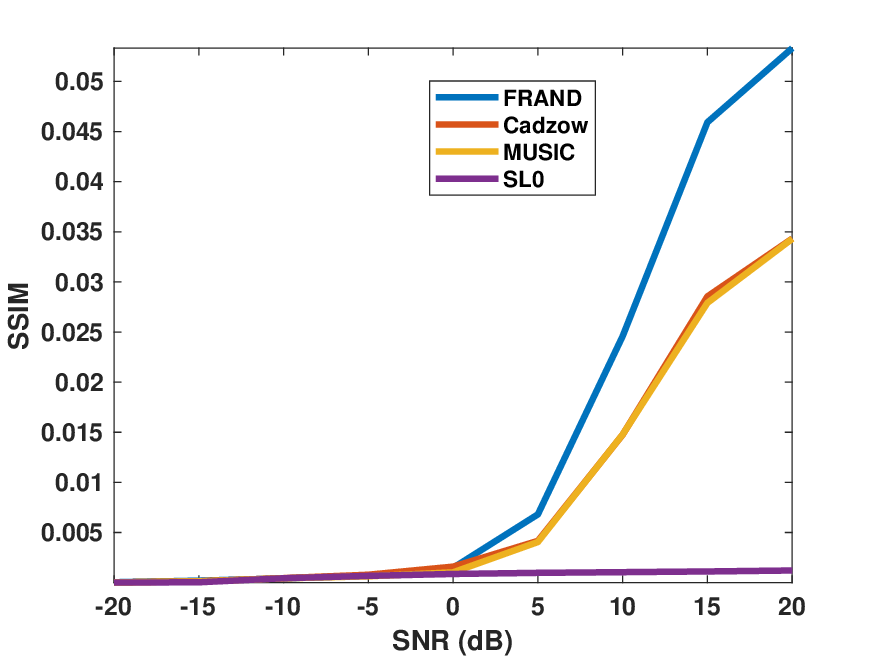}
\caption{Comparison of FRAND with MUSIC, Cadzow, and $\mathrm{SL}_{0}$ in term of SSIM criterion and  SNRs.}
\label{fig8}
\end{figure}

In the next simulation, the execution speeds of different algorithms are compared in Fig~\ref{fig9} for 100  to 500 samples.  Although the execution time of FRAND is higher than the others, it can achieve better resolution for ISAR imaging of small targets. It should be pointed out that the execution time of FRAND without implementing the 2D-ADMM algorithm is 7654 seconds for 200 samples, which is reduced to 1.1 seconds by incorporating 2D-ADMM.
\begin{figure}
\includegraphics[angle=0,width=0.65\textwidth]{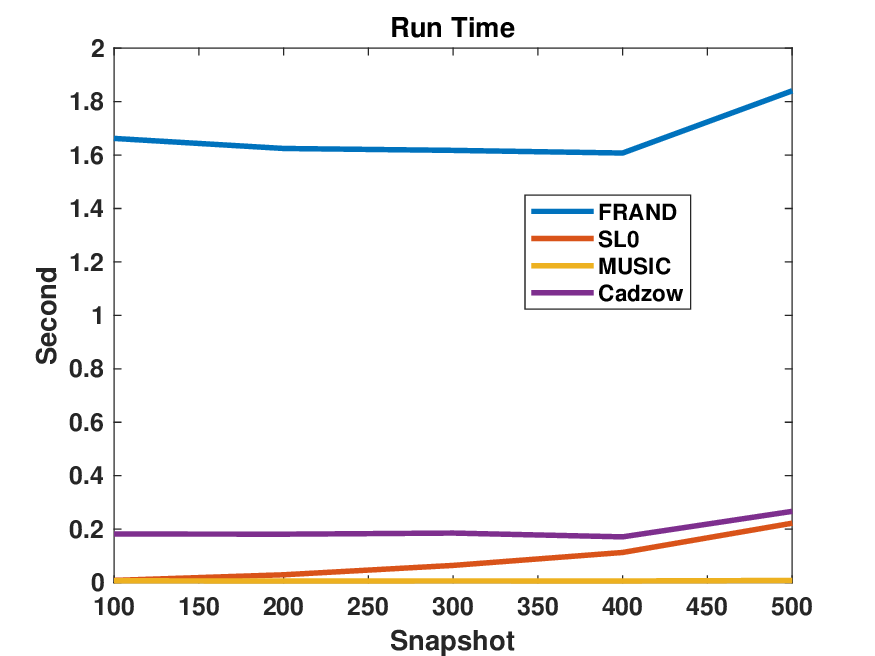}
\caption{Comparison of running time of FRAND, MUSIC, Cadzow, and $\mathrm{SL}_{0}$ for available samples from 100 to 500.}
\label{fig9}
\end{figure}
\section{Conclusion}
The  FRAND method  was proposed based on weighted atomic norm minimization and the 2D-ADMM algorithm to reduce the noise effect and enhance ISAR image resolution in both range and azimuth directions. This super-resolution algorithm was developed based on retrieving more sparse scatterer points of a target in sparse apertures. In this method, the atomic norm of the received signal was written as a linear combination of a set of atoms, which were weighted by a matrix. Then, the 2D-ADMM algorithm was implemented for this problem in order to reduce the  number of computations and running time. Simulation results showed the superiority of FRAND over the MUSIC, Cadzow, and $\mathrm{SL}_{0}$ algorithms at different SNRs in terms of the MSE, PSNR, and SSIM criteria. The results demonstrate the improved efficiency of the proposed method in reducing  noise effect and thus increasing the ISAR image resolution.

\section*{Author Contributions}

\paragraph*{Formal analysis:}
\label{Mohammad Roueinfar.}
{Mohammad Roueinfar.}  

\paragraph*{Methodology:}
\label{Methodology}
 {Mohammad Roueinfar., Mohammad Hossein Kahaei.}

\paragraph*{Supervision:}
\label{Supervision:}
{Mohammad Hossein Kahaei.}

\paragraph*{Validation:}
\label{Validation:}
{Mohammad Hossein Kahaei.}

\paragraph*{Writing – original draft:}
\label{Writing – original draft:}
{ Mohammad Roueinfar.}

\paragraph*{Writing – review and editing:}
\label{Writing – review and editing:}
{ Mohammad Hossein Kahaei.}

\nolinenumbers

%
%
%

\end{document}